\documentclass[a4paper,11pt]{extarticle}
\usepackage[utf8]{inputenc}

\usepackage{geometry}
\usepackage{xcolor}
\usepackage{amsmath, bm}
\usepackage{amsfonts}
\usepackage{graphicx}
\usepackage{makecell}
\usepackage{framed}
\usepackage{multirow}
\usepackage{empheq}
\usepackage{appendix}
\usepackage{caption}
\usepackage{comment}
\DeclareMathOperator*{\argmin}{\arg\!\min}
\title{Learning fast, accurate, and stable closures of a kinetic theory of an active fluid}
\author
{Suryanarayana Maddu$^{\textrm{a},\ast}$ , Scott Weady$^\textrm{a}$ , Michael J. Shelley$^{\textrm{a,b}}$
	\\ 
	\footnotesize{$^{\textrm{a}}$ \footnotesize Center for Computational Biology, Flatiron Institute, Simons Foundation, New York, NY, 10010, USA}\\
	\footnotesize{$^{\textrm{b}}$ \footnotesize Courant Institute of Mathematical Sciences, New York University, New York, NY, 10012, USA}
	\\
	\footnotesize{$^\ast$Correspondence to: smaddu@flatironinstitute.org}
}

\date{\vspace{-2em}}
\begin{document}

\maketitle

\begin{abstract}
    Important classes of active matter systems can be modeled using kinetic theories. However, kinetic theories can be high dimensional and challenging to simulate. Reduced-order representations based on tracking only low-order moments of the kinetic model serve as an efficient alternative, but typically require closure assumptions to model unrepresented higher-order moments. 
    In this study, we present a learning framework based on neural networks that exploits rotational symmetries in the closure terms to learn accurate closure models directly from kinetic simulations. The data-driven closures demonstrate excellent \textit{a-priori} predictions comparable to the state-of-the-art Bingham closure. We provide a systematic comparison between different neural network architectures and demonstrate that nonlocal effects can be safely ignored to model the closure terms. We develop an active learning strategy that enables accurate prediction of the closure terms across the entire parameter space using a single neural network without the need for retraining. We also propose a data-efficient training procedure based on time-stepping constraints and a differentiable pseudo-spectral solver, which enables the learning of stable closures suitable for \textit{a-posteriori} inference. The coarse-grained simulations equipped with data-driven closure models faithfully reproduce the mean velocity statistics, scalar order parameters, and velocity power spectra observed in simulations of the kinetic theory. Our differentiable framework also facilitates the estimation of parameters in coarse-grained descriptions conditioned on data.
\end{abstract}

\section{Introduction}

The field of active matter has found exciting applications in many areas such as biophysics, synthetic chemistry, and material science \cite{ramaswamy2010mechanics, marchetti2013hydrodynamics, saintillan2013active}. Active matter systems consist of collections of agents, such as particles or macromolecules, that convert chemical energy into mechanical work, creating active forces or stresses that are communicated across the system via direct steric interactions, cross-linking proteins, or through long-ranged hydrodynamic interactions. The consequent coordination can lead to complex spatiotemporal dynamics. A central question in these systems is the relation between the microscopic interactions on the scale of individual particles and the emerging self-organization and collective dynamics at the scale of the system \cite{vicsek2012collective, saintillan2013active}. In this study, we focus on active suspensions, in which the active particles are suspended in a viscous fluid and long-ranged hydrodynamic interactions, alongside steric interactions, are important. A few prominent experimental realizations include suspensions of swimming bacteria and \textit{in vitro} mixtures of cytoskeletal filaments and motor proteins. These systems exhibit flows characterized by large-scale vortices and jets with sizes and speed much greater than those associated with the individual swimmer or motor \cite{dombrowski2004self, dunkel2013fluid}. Though these spontaneous flows exhibit turbulent characteristics, energy input is due to its microscopic components.
Hence, in active matter systems, the flows generated are autonomous, and the emerging pattern of energy injection is self-organized \cite{alert2022active}.

\indent Many computational and mathematical approaches have been used to investigate active matter systems, ranging from large-scale discrete particle simulations to continuum formulations. Discrete simulations explicitly model the dynamics of active particles and their hydrodynamic and steric interactions. These models have proven invaluable and are instrumental to achieving qualitative agreements with experimental observations. However, due to their discrete nature, such descriptions cannot be easily analyzed mathematically \cite{saintillan2013active} and remain constrained in the number of particles simulated. Several continuum models have been proposed for studying active suspensions based on the extension of the theory of passive liquid crystals to include activity \cite{simha2002hydrodynamic}, and phenomenological models constructed based on symmetries to include steric interactions \cite{toner1998flocks, dunkel2013fluid}. Alternate approaches based on coupling the microscopic and macroscopic scales include kinetic theories where particle positions and their conformations, such as particle orientation, are represented through a distribution function and evolved according to the Smoluchowski equation coupled to the equations of activity-driven hydrodynamics. Even though kinetic theories involve coefficients that are well grounded in microscopic details and are amenable to analytical treatment, they are not immune from computational challenges. The particle distribution function depends on both particle position and orientation leading to $2d\!-\!1$ degrees of freedom. In dense suspensions with strong alignment interactions, the orientation field needs to be resolved with very high accuracy, which prohibitively increases the cost of kinetic simulations, even in two dimensions. To circumvent this, approximate coarse-grained models that only track the first few orientational moments of the distribution function have been proposed. However, evolution equations for the moments are not closed as they depend on unrepresented higher-order moments of the  distribution function. Several closure models have been proposed in the context of liquid crystals and particle suspensions with approximations based upon weak or strong flow or near isotropy \cite{ezhilan2013instabilities, saintillan2013active}, or by interpolating between such states  \cite{hinch1976constitutive}. In addition, accurate and efficient means exist for both  polar \cite{weady2022thermodynamically} and apolar suspensions \cite{weady2022fast, gao2017analytical} that rely on modeling the particle distribution function by a Bingham distribution on the unit sphere. Most of these approximations either involve assuming quasi-equilibrium approximation of the distribution function or closures that apply to specific flow regimes \cite{saintillan2013active}.

In this paper, we propose a data-driven framework for learning closure models directly from data generated from kinetic theory. We draw inspiration from fluid dynamics applications where machine learning has been used to model discrepancies between a prescribed phenomenological closure and high-fidelity direct numerical simulations (DNS)\cite{brunton2020machine}. 
Most of the studies involve the use of supervised learning techniques where a neural network is employed to learn corrective dynamics using DNS data as the target. Notable approaches include incorporating Galilean invariance into the Reynolds stress tensor predictions using neural networks (NN) \cite{ling2016machine,ling2016reynolds}, learning source terms from coarsely resolved quantities through NNs \cite{maulik2019subgrid}, and discovery of closed-form equations for coarse-grained ocean models using high-resolution data \cite{zanna2020data}. Similarly, the study \cite{parish2016paradigm} developed a field inversion and ML modeling framework to learn corrections to the Spalart–Allmaras RANS model \cite{singh2017machine}. To address the issue of compounding modeling errors during inference, ideas based on multi-agent reinforcement learning were used to locally adapt the coefficients of the eddy-viscosity closure model, with the objective to reproduce the energy spectrum predicted by DNS. In contrast to supervised learning, training in this case was not performed on a database of reference data, but through time integration of the parametric model and penalizing unstable models. While the latter approach provides stable \textit{a-posteriori} predictions, its exploratory nature makes it computationally expensive. In this context, integrating neural network models with differential numerical solvers has enhanced the temporal stability of inferred closures, yielding significant improvements in long-term {\it a posteriori} statistics for a range of unsteady simulations \cite{list2022learned, sirignano2020dpm, kochkov2021machine, macart2021embedded}, while maintaining computational efficiency.\\
\indent By drawing parallels between the closure problem arising in active fluids and classical closure modeling in fluid dynamics, we successfully adapt ideas from ML-based closure modeling to develop data-driven closures for active matter systems. To begin, we generate high-fidelity data through kinetic theory, which serves as reference data for learning empirical closures models or comparing against. This is similar to the use of DNS data as  ground truth in developing data-driven closures in fluid dynamics. Next, we identify the rotational symmetries arising in closure terms of the coarse-grained description of dense active nematics and leverage a second-order tensor-valued isotropic representation to learn an accurate and stable approximation of the closure terms. To overcome the distinction between \textit{a priori} and \textit{a posteriori} evaluation, we propose a strategy based on 
discrete time-stepping constraints that embed temporal stability into the learned closures.\\
\indent The main contributions of this work are as follows. We focus on the coarse-grained description of the active apolar suspension derived from tracking only the zeroth (concentration) and second (orientation tensor) moment of the particle distribution function.
By recognizing the isotropic nature of the closure terms, we utilize isotropic second-order tensor representation to model them. The proposed learning formulation incorporates invariant input features, enabling a rotationally invariant representation of the learned closures, resulting in improved accuracy and data efficiency compared to approaches without encoded symmetry. We provide a quantitative comparison between different neural architectures with varying receptive fields and mathematical representation, and also compare the data-driven approximations with commonly used closure models. To ensure stable \textit{a-posteriori} prediction, we propose an optimization formulation based on model predictive control (MPC) with time-stepping constraints, ensuring temporal stability of the learned closures. In the last section, we assess the accuracy of the closure models for solving inverse problems, specifically inferring parameters of coarse-grained descriptions using data from kinetic theory.

\section{Active suspension model}\label{section:active_model}
We are interested in studying the dynamics of $N$ active and immersed particles of length $\ell$ and thickness $b$ (aspect ratio $= \ell/b \gg 1$) in a volume $V$ assumed to be a cube of linear dimension $L = \vert V\vert^{1/3}$ \cite{ezhilan2013instabilities,gao2017analytical}. The dynamics of a concentrated suspension of such particles is modeled using kinetic theory, wherein the particle configuration is described by means of a continuum distribution function $\Psi(\mathbf{x}, \mathbf{p}, t)$ of the center of mass $\mathbf{x}$ and orientation vector $\mathbf{p}$ at time $t$. The zeroth, first, and second moments of the distribution function with respect to $\mathbf{p}$ correspond to the concentration field $c = \langle 1 \rangle$, polarity field $\mathbf{n} = \langle\mathbf{p}\rangle/c$, and nematic order parameter $\mathbf{Q} = \langle \mathbf{p}\mathbf{p}\rangle/c$, respectively, where $\langle f \rangle = \int_{|\mathbf{p}|=1} f \Psi ~ d\mathbf{p}$ denotes an orientational moment.
The evolution of the distribution function is governed by the Smoluchowski equation, reflecting the conservation of particle number, and is given by
\begin{equation}\label{fokker_planck}
   \frac{\partial \Psi}{\partial t} + \nabla_{\mathbf{x}} \cdot \left( \dot{\mathbf{x}}\Psi \right) + \nabla_{\mathbf{p}} \cdot \left( \dot{\mathbf{p}}\Psi \right) = 0,
\end{equation}
\noindent where the conformational fluxes $\dot{\mathbf{{x}}}$ and $\dot{\mathbf{{p}}}$ are obtained from the dynamics of a single particle in a background flow $\mathbf{u}(\mathbf{x},t)$. The operator $\nabla_{\mathbf{p}} = \left(\mathbf{I} - \mathbf{pp} \right) \cdot (\partial/\partial \mathbf{p})$ denotes the gradient operator on the unit sphere of orientations. For a concentrated suspension, the conformational fluxes are
\begin{align}
    \dot{\mathbf{{x}}} &= \mathbf{u} - d_T \nabla_\mathbf{x} \log \Psi, \label{eq:com} \\
    \dot{\mathbf{{p}}} &= (\mathbf{I} - \mathbf{pp}) \cdot \left( \nabla \mathbf{u} + 2 \zeta \mathbf{D}\right) \cdot \mathbf{p} -d_R \nabla_\mathbf{p} \log \Psi. \label{eq:pol}
\end{align} 
Here $d_T$ and $d_R$ are dimensionless translational and rotational diffusion constants, $\zeta$ is the strength of particle alignment through steric interactions, and $\mathbf{D} = \langle\mathbf{p}\mathbf{p}\rangle$ is the second-moment tensor. Equation~(\ref{eq:com}) describes the particles being advected by the local mean-field velocity $\mathbf{u}$ and diffusing isotropically. Similarly, in Eq.~(\ref{eq:pol}), the angular flux velocity of the particle is dictated by the rotation of a slender rodlike particle by the mean-field quantity $\nabla \mathbf{u}+2\zeta \mathbf{D}$, as well as rotational diffusion. The Smoluchowski equation is coupled to the Stokes flow as
\begin{align}
\quad & -\Delta \mathbf{u} + \nabla P = \nabla \cdot \mathbf{\Sigma}, \:\:\: \nabla \cdot \mathbf{u} = 0 \label{eq:stokes},  \\
\color{black}\mathbf{\Sigma} &= \alpha \mathbf{D} +  \beta \mathbf{S\!:\!E} \color{black} - 2 \zeta \beta \left(\mathbf{D} \cdot \mathbf{D} - \mathbf{S\!:\!D}\right). \label{eq:stress_tensor}
\end{align}
Here $P(\mathbf{x},t)$ is the fluid pressure, $\alpha$ is the dimensionless active dipole strength, $\beta$ characterizes the particle density, $\mathbf{E} = [\nabla \mathbf{u} + \nabla \mathbf{u}^{\top}] /2$ is the symmetric rate-of-strain tensor, and $\mathbf{S} = \langle\mathbf{p}\mathbf{p}\mathbf{p}\mathbf{p}\rangle$ is the fourth-moment tensor. The stress tensor $\mathbf{\Sigma}$ in Eq.~(\ref{eq:stress_tensor}) is written as the sum of three contributions arising from the active dipole, particle rigidity, and stress arising from local steric torques, respectively.
The system of equations~(\ref{fokker_planck})-(\ref{eq:stokes}) form a closed system which is referred to as the kinetic theory. 

To circumvent the computational complexity arising from the high dimensionality of the kinetic model, we evolve only certain moments of the distribution function $\Psi$. Specifically, for apolar suspensions, such as the one considered here, we evolve only the zeroth and second moments $c$ and $\mathbf{D}$. By taking moments of the Smoluchowski equation~(\ref{fokker_planck}), we can derive:
\begin{align}
\color{black}\frac{\partial c}{\partial t} + \mathbf{u} \cdot \nabla c &= \color{black}d_T \Delta c,  \label{coarse_grain_c}\\
\color{black}\mathbf{D}^{\nabla} + 2 \mathbf{S\!:\!E} &= \color{black}4\zeta \left( \mathbf{D\!\cdot\! D} - \mathbf{S\!:\!D} \right) + d_T \Delta \mathbf{D} - 2dd_R \left( \mathbf{D} - \frac{c}{d} \mathbf{I}\right). \label{coarse_grain_D}
\end{align}
Here $\mathbf{D}^{\nabla} = \partial \mathbf{D}/\partial t + \mathbf{u} \cdot \nabla \mathbf{D} - \left( \nabla \mathbf{u} \cdot  \mathbf{D} + \mathbf{D} \cdot \nabla \mathbf{u}^\top \right) $ is the upper-convected time derivative and $d$ is the spatial dimension. Equations~(\ref{coarse_grain_c})-(\ref{coarse_grain_D}) involve the higher fourth-moment of the distribution function $\mathbf{S}$. Hence, Eqs.~(\ref{eq:stokes})-(\ref{coarse_grain_D}) are not closed and one needs to employ a \textit{closure model} to approximate terms such as $\mathbf{S\!:\!D}$ and $\mathbf{S\!:\!E}$ in terms of $c$ and $\mathbf{D}$. 
We also observe that Eqs.~$(\ref{eq:stokes})$ and $(\ref{coarse_grain_D})$ depend only on the symmetric second-order tensor $\mathbf{S\!:\!T}$ with $\mathbf{T} := \mathbf{E} + 2\zeta \mathbf{D}$. In our study, we concern ourselves with the representation and learning of this tensorial quantity $\mathbf{S\!:\!T}$ from kinetic simulations. 




\section{Closure representation and the learning problem}\label{sec:invariant_rep}
The term $\mathbf{S\!:\!T}$  described in the previous section has the following rotational symmetry
\begin{equation*}
    \mathbf{\mathbf{S}^{'}\!:\!\mathbf{T}^{'}}=   \mathbf{\Omega} \left( \mathbf{S\!:\!T} \right) \mathbf{\Omega}^\top,
\end{equation*}
where $\mathbf{T}^{'} = \mathbf{\Omega} \mathbf{T} \mathbf{\Omega}^\top$. To put it simply, if we rotate the tensor arguments $(\mathbf{S}, \mathbf{D},\mathbf{E})$ of the contraction, the resulting tensor contraction $\mathbf{S\!:\!T}$ should also be rotated by the same angle. Since $\mathbf{S\!:\!T}$ is computed through a linear combination of $\mathbf{S\!:\!D}$ and $\mathbf{S\!:\!E}$, as a consequence, both $\mathbf{S\!:\!D}$ and $\mathbf{S\!:\!E}$ tensors also adhere to the aforementioned relation.  Any tensor-valued function that obeys the above rotational equivariance is termed an isotropic tensor-valued function. More formally, a second-order tensor valued function $\mathbf{F}(\mathbf{A}_1,...,\mathbf{A}_M)$ is isotropic if
\begin{equation}\label{eq:tensor_rep}
    \mathbf{F}(\mathbf{\Omega} \mathbf{A}_1 \mathbf{\Omega}^\top,...,\mathbf{\Omega} \mathbf{A}_M \mathbf{\Omega}^\top) = \mathbf{\Omega} \: \mathbf{F}(\mathbf{A}_1,...,\mathbf{A}_M)\: \mathbf{\Omega}^\top, \quad \forall \: \mathbf{\Omega}  \in \text{O}(d) 
\end{equation}
where $\{ \mathbf{A}_i \}_{i=1,2,...,M}$ are symmetric tensors and $\textrm{O}(d)$ is the full orthogonal group in $d$-dimensions. In general, a symmetric tensor-valued isotropic function $\mathbf{F}(\mathbf{A}_1,\ldots,\mathbf{A}_M)$ can be expanded as
\begin{equation}\label{eq:tensor_expansion}
    \mathbf{F}(\mathbf{A}_1,...,\mathbf{A}_M) = \sum_{m=1}^{M} \varphi_m(\mathcal{I}_s) \mathbf{F}_m,
\end{equation}
\noindent where $\mathbf{F}_m$ are form-invariant symmetric tensor-valued isotropic functions, referred to as generators, and $\varphi_m$ are scalar-valued isotropic functions of the invariants $\mathcal{I}_s$ of the functional basis of $\{\mathbf{A}_i\}_{i=1,...,M}$ \cite{korsgaard1990representation}. By representing the tensor-valued function using a set of fundamental tensors, known as an integrity basis, we can ensure tensor isotropy, as described in Eq.~(\ref{eq:tensor_rep}). Such a representation has been used in classical fluid turbulence closures to model the Reynolds stress anisotropy tensor, where the anisotropic tensor is expanded in the integrity basis of the strain rate tensor and the mean vorticity tensor \cite{pope1975more, ling2016machine}. For the specific case of dense active apolar suspensions, we are interested in the representation of the tensor-valued function $\mathbf{F(S,D,E)} = \mathbf{S\!:\!E} + 2\zeta \mathbf{S\!:\!D}$. Under the assumption of uniform concentration $c \equiv 1$ and that the fourth-order tensor $\mathbf{S}$ depends only on the known lower order moment $\mathbf{D}$, the quantity $\mathbf{S\!:\!D}$ is a tensor-valued function of $\mathbf{D}$ only, and $\mathbf{S\!:\!E}$ is a tensor-valued function with tensor arguments $\left(\mathbf{D,E}\right)$.

Let us first consider the representation of the tensor quantity $\mathbf{S\!:\!D}$ in two dimensions. The independent invariant quantities associated with the orientation tensor $\mathbf{D}$ are $\mathcal{I}_s(\mathbf{D}) = \{ \textrm{tr}\mathbf{D},  \textrm{tr}\mathbf{D}^2 \}$ and the corresponding tensor/generator basis is $\mathbf{F}_m = \{ \mathbf{I},  \mathbf{D} \}$. Therefore, under the isotropic representation described in equation~(\ref{eq:tensor_expansion}), we can write
\begin{equation}\label{eq:SD}
    \mathbf{S\!:\!D} = \gamma_0 \mathbf{I} + \gamma_1  \mathbf{D},
\end{equation}
where $\gamma_0,\gamma_1$ are scalar-valued isotropic functions of the invariants $\mathcal{I}_s(\mathbf{D})$. 
In a similar fashion, the tensor quantity $\mathbf{S\!:\!E}$ can be written in the expanded form,
\begin{equation}\label{eq:SE}
     \mathbf{S\!:\!E} = \kappa_0  \mathbf{I} + \kappa_1 \mathbf{D} + \kappa_2 \mathbf{E},
\end{equation}
where $\{ \kappa_i \}_{i=0,1,2}$ are scalar isotropic functions of the invariants $\mathcal{I}_s(\mathbf{D,E}) = \{ \textrm{tr}\mathbf{D},\textrm{tr}\mathbf{E}, \textrm{tr}\mathbf{D}^2,  \textrm{tr}\mathbf{E}^2,  \textrm{tr}\mathbf{DE}\}$. On the account of the incompressibility condition $\nabla \! \cdot \!\mathbf{u} = 0$ and the assumption of uniform concentration, we can safely omit the invariants $\textrm{tr}\mathbf{D}$ and $\textrm{tr} \mathbf{E}$ from the set $\mathcal{I}_s(\mathbf{D},\mathbf{E})$. The detailed derivation of the representation is discussed in Appendix \ref{derivation}. 
Combining equations~(\ref{eq:SD}) and (\ref{eq:SE}) the closure tensor quantity $\mathbf{S\!:\!T}$ can be written in the expanded form as,
\begin{equation}\label{isotropic_rep}
    \mathbf{F}(\mathbf{D,E}) = \mathbf{S\!:\!T} = \left( \kappa_0 + 2 \zeta \gamma_0 \right) \mathbf{I} + \left(\kappa_1 + 2\zeta \gamma_1\right) \mathbf{D} + \kappa_2 \mathbf{E}.
\end{equation}
It is straightforward to show that the above representation satisfies the tensor isotropic relation discussed in equation~(\ref{eq:tensor_rep}), i.e.
\begin{align*}
    \mathbf{F}(\mathbf{\Omega D\Omega }^\top,\mathbf{\Omega E\Omega }^\top) &= \left( \kappa_0 + 2 \zeta \gamma_0 \right) \mathbf{I} + \left(\kappa_1 + 2\zeta \gamma_1\right) \mathbf{\Omega D\Omega }^\top + \kappa_2 \mathbf{\Omega E\Omega }^\top.\\
    &= \mathbf{\Omega} \left(\left( \kappa_0 + 2 \zeta \gamma_0 \right) \mathbf{I} + \left(\kappa_1 + 2\zeta \gamma_1\right) \mathbf{D} + \kappa_2 \mathbf{E} \right) \mathbf{\Omega}^\top = \mathbf{\Omega} \mathbf{F}(\mathbf{D,E})\mathbf{\Omega}^\top.
\end{align*}
\noindent The isotropic representation provides a relation on how the tensor contraction $\mathbf{S}\!\!:\!\!\mathbf{T}$ rotates if its input arguments are transformed by a rotation matrix $\mathbf{\Omega} \in \textrm{O}(2)$. More importantly, the relation~(\ref{isotropic_rep}) provides a strategy for learning closure representations that by construction satisfy properties of tensor-valued isotropic functions. In the next subsection, we discuss how a learning problem can be formulated based on the isotropic representation, where we try to learn the nonlinear dependency of the coefficients $(\gamma_0,\gamma_1,\kappa_0,\kappa_1,\kappa_2)$ on the invariant quantities $\mathcal{I}_s(\mathbf{D},\mathbf{E})$.


\subsection{Optimization formulation}

\noindent We use neural networks to learn the nonlinear dependency between the scalar-valued isotropic function coefficients $(\gamma,\kappa)$ and the invariant quantities $\mathcal{I}_s(\mathbf{D},\mathbf{E})$. Thus inferred coefficients are used to reconstruct the closure terms using the expansion described in Eq.~(\ref{isotropic_rep}). To achieve this, we leverage the universal function approximation properties of neural networks within a supervised learning framework. In this setting, the network parameters $(\theta)$ are optimized to learn a nonlinear map between an input vector and the predefined target vector (output). 
The closure learning problem within a supervised setting can be written as follows,
\begin{equation*}
    \widetilde{\theta} = \argmin_{\theta} \int_{\Omega_d} \left(\big \Vert \mathbf{S\!:\!T}(\mathbf{x}) -  (\kappa_ {0}(\theta) + 2 \gamma_0(\theta))\mathbf{I} - (\kappa_{1}(\theta) + 2 \gamma_{1}(\theta) )\mathbf{D}(\mathbf{x}) - \kappa_2(\theta) \mathbf{E}(\mathbf{x})  \big \Vert_F^2 \right) d \mathbf{x},
\end{equation*}
\noindent where the neural network, $\mathcal{N}_\theta:   \big\{ \textrm{tr}\mathbf{D}^2,\textrm{tr}\mathbf{E}^2,\textrm{tr}\mathbf{DE} \big \} \rightarrow \big\{ \gamma_0,\gamma_1,\kappa_0, \kappa_1,\kappa_2 \big \} $ with parameters $\theta$, is optimized to learn the map between the invariant feature space and the coefficients of the isotropic representation in Eq.~(\ref{isotropic_rep}), and $\Omega_d$ is the spatial domain of dimension $d$. 
The reference data for the tensor quantity $\mathbf{S\!:\!T}$ is generated from numerically solving the kinetic theory described in Eqs.~(\ref{fokker_planck})-(\ref{eq:stokes}). As a baseline comparison with the isotropic representation, we learn a direct nonlinear map between the  components of tensor function arguments $\left(\mathbf{D,E}\right)$ and the closure tensor $\mathbf{S\!:\!T}$ by solving the optimization problem
\begin{equation*}
    \widetilde{\theta} = \argmin_{\theta} \int_{\Omega_d} \left(\big \Vert \mathbf{S\!:\!T}(\mathbf{x}) - \widetilde{\mathbf{S\!:\!T}}_\theta(\mathbf{x}) \big \Vert_F^2 \right)d\mathbf{x}.
\end{equation*}
\noindent Here, the neural network $\mathcal{N}_\theta$ learns the map $\mathcal{N}_\theta: \{\mathbf{D,E} \} \rightarrow  \{\widetilde{\mathbf{S\!:\!T}_\theta}\}$. This straightforward nonlinear map between tensor components is not constrained to satisfy the isotropic function property of the tensor-valued function $\mathbf{S\!:\!T}$. Hence, it provides an ideal baseline for evaluating the impact of rotational invariance on the accuracy and extrapolation properties of the inferred closure models. All of this relies on the approximation capabilities of neural networks $\mathcal{N}_\theta$. In the next section we provide a brief description of neural networks and highlight different network architectures with varying receptive fields and internal mathematical representations.
\begin{figure}[htbp]
  \centering
  \includegraphics[width=7in]{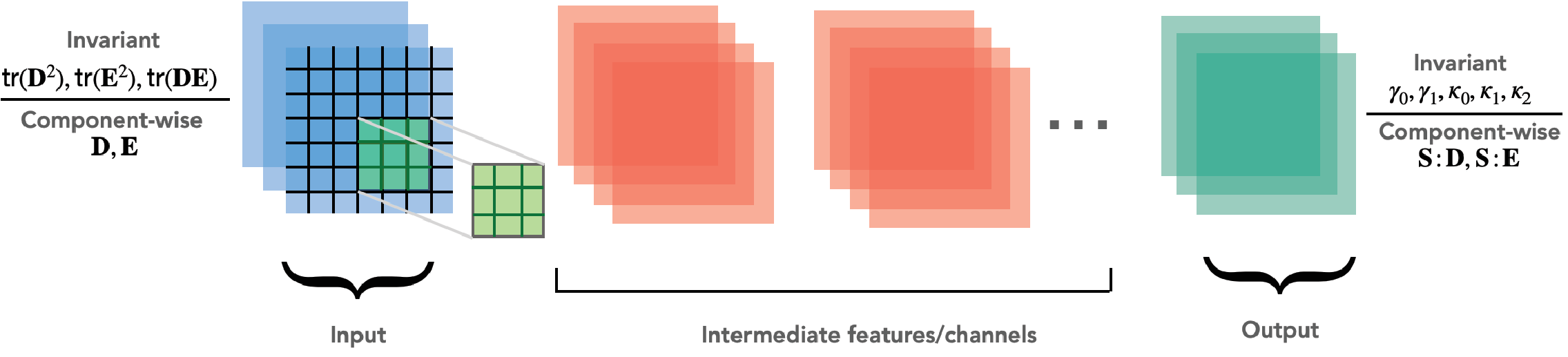}
  \caption{\textbf{Learning based on isotropic tensor representation v/s Component-wise mapping:} For the isotropic representation of closure terms, the input to the neural network (CNN discussed here) are rotationally invariant scalar quantities of the orientation tensor $\mathbf{D}$ and the strain-rate $\mathbf{E}$. The outputs are the coefficients $\{\gamma_0, \gamma_1, \kappa_0, \kappa_1, \kappa_2\}$ that are used to compose the closure terms through the relation $\mathbf{S\!:\!T} =  (\kappa_0 + 2 \zeta \gamma_0) \mathbf{I} + (\kappa_1 + 2 \zeta \gamma_1) \mathbf{D} + \kappa_2 \mathbf{E}$. The input to the neural network is the tensor components of the orientation tensor $\mathbf{D}$ and the strain-rate $\mathbf{E}$. The outputs are the tensor components of the tensors $\mathbf{S\!:\!D}$ and $\mathbf{S\!:\!E}$ which are then composed to approximate the closure terms, i.e. $\mathbf{S\!:\!T}=\mathbf{S\!:\!E}+2\zeta\mathbf{S\!:\!D}$. The gridded $3\times3$ box corresponds to the receptive field of the CNN. } 
  \label{fig:comp_rep}
\end{figure}
\subsection{Neural network architectures}\label{sec:neural_nets}

In general, a Neural Network (NN) is a multivariate compound function which contains a number of free parameters, called weights and biases, that can be learned. It maps an input vector to an output vector via successive linear (matrix multiplication) and non-linear operations. The set of real-valued $n$-layer neural network with input $\mathbf{z}$ can be written as a function,
\begin{align}\label{DNN}
&\mathcal{N}_\theta := \{f:\mathbb{R}^{\text{inp}} \rightarrow \mathbb{R}^\text{out} \}, \nonumber\\
    &f(\mathbf{z}) = \mathcal{T}_n \left(\varrho \left(\mathcal{T}_{n-1}\left(\varrho( \hdots \mathcal{T}_1\left( \varrho(\mathcal{T}_0(\mathbf{z}))\right) \hdots) \right)\right)\right),
\end{align}
\noindent where $\varrho:\mathbb{R} \rightarrow \mathbb{R}$ is an element-wise nonlinear activation function and $\mathcal{T}_k(\mathbf{z}) = \mathbf{W}_k \mathbf{z} + \mathbf{b}_k$ is the affine linear map with trainable parameters $(\theta)$ in the form of weights and biases, i.e. $\theta = \{ \textbf{W}_k, \textbf{b}_k \}_{k=0,1,\ldots , n}$. The number of hidden layers (all layers excluding the input and output layer) determines the depth of the neural network, and the number of nodes per layer are important hyperparameters that determine the expressibility and training requirements of the network. The way nodes are connected and information is passed along is encoded in the matrices $\mathbf{W}_k$. For instance, in a fully connected neural network (MLP) the matrices $\mathbf{W}_k$ are dense and in Convolutional Neural Networks (CNN) the matrices are usually sparse and circulant \cite{beck2019deep}. 
\begin{figure}[htbp]
  \centering
  \includegraphics[width=3.in]{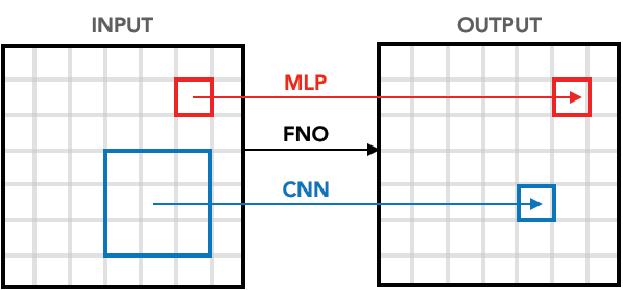}
  \caption{\textbf{Receptive field and response field of different neural architectures:} The MLP architecture learns a map between the input attributes at a point in the spatial domain to output attributes at the same point. CNN architectures rely on local patch-wise convolutions to map a receptive field in the input domain to a point in the output domain. Fourier neural operators (FNO) rely on convolutions in Fourier space (global convolutions) to learn the mapping, resulting in a receptive field that spans the entire spatial domain. In the context of the closure problem, the input channel corresponds to a spatial grid containing the tensor components or its invariants.} 
  \label{fig:receptive_field}
\end{figure}

In CNN architectures, the input attributes in the receptive field are mapped to output quantities through convolutions based on learnable filters. Such patch-wise learning with filters is used for automatic feature extraction and thus CNN architecture finds extensive applications in image recognition and object detection \cite{lecun1989handwritten, lecun1995convolutional}. Though CNNs are usually used for feature extraction, in this study we use them to learn the nonlinear map associated with the closure. At the level of mathematical operations, CNNs are closely related to MLPs, except the global matrix multiplication is replaced by a local, multi-dimensional convolution filtering operator. The affine linear transformation in the $k^{th}$ layer of CNN is,
\begin{equation}\label{eq:CNN_feature}
\mathcal{T}_k(\mathbf{z}) = \mathbf{W}_k  \mathbf{z} + \mathbf{b}_k,
\end{equation} 
where $\mathbf{W}_k$ is a sparse circulant matrix that represents the convolution and depends on the size of the filter, and $\mathbf{b}_k$ is the bias term. For a given feature, the same local convolution filter is applied to the whole input field to extract hierarchical features. 
Similarly, if we replace or append the local convolutions with global convolutions through spatial Fourier transforms, we arrive at Fourier Neural Operators (FNO) \cite{li2020fourier}. The affine linear transformation in the $k^{th}$ layer of FNO is,
\begin{equation}\label{FNO}
\mathcal{T}_{k}(\mathbf{z}) =  \mathbf{W}_k  \mathbf{z} + \mathbf{b}_k + \mathcal{F}^{-1} ( \mathbf{R}_{k} \mathcal{F}\left( \mathbf{z} )\right),
\end{equation} 
where $\theta = \{ \textbf{W}_k, \, \textbf{R}_k,\, \textbf{b}_k \}$ are the trainable weights and biases. The operators $\mathcal{F}$ and $\mathcal{F}^{-1}$ are the forward and inverse discrete Fourier transforms, respectively. FNOs have been used to learn PDE solution operators wherein non-local effects are captured through global convolutions \cite{li2020fourier, kovachki2021neural}. The single layer operations of CNN and FNO are then used to compose the multivariate compound functions as shown in equation~(\ref{DNN}).

One important distinction between different neural architectures is how the input features are processed as shown in Figure~\ref{fig:receptive_field}. In MLP, the input features are usually transformed to vectors before feeding into the neural network whereas in CNN and FNO the original multidimensional structure of the input features is preserved. Also, MLPs learn a point-wise map between input and target features, however in CNN the receptive field is a local multidimensional neighborhood around the point where the output prediction is desired. As depicted in Figure~\ref{fig:receptive_field}, the receptive field in FNOs encompasses the entire spatial domain as it relies on global Fourier convolutions. The MLP neural architecture is agnostic to the grid resolution and domain-size as it involves point-wise mapping between input and output quantities.




\section{Numerical results}

In this section, we conduct \textit{a-priori} and \textit{a-posteriori} analyses of the closure approximations discussed in the previous sections. To generate the reference data from kinetic theory, we use a pseudo-spectral discretization of Eqs. (\ref{fokker_planck})-(\ref{eq:stokes}) with the 2/3 anti-aliasing rule where Fourier differentiation is used to evaluate the derivatives with respect to space and particle orientation. We use a second-order implicit-explicit backward differentiation time-stepping scheme (SBDF2), where the linear terms are handled implicitly and the nonlinear terms explicitly with time-step $\Delta t = 0.0004$.  The numerical simulations are performed in a periodic square domain of length 
$L=10$ with $256^2$ spatial modes and $256$ orientational modes. We initialize $\mathbf{D}$ with a plane-wave perturbation about the isotropic state $\mathbf{D}_0 = \mathbf{I}/2$ such that that $\text{tr}\mathbf{D} = 1$ and take the concentration to be uniform so that  $c(\mathbf{x},t)\equiv 1 $. In concentrated suspensions, the isotropic-to-nematic transition occurs when the dimensionless parameter $\zeta$ is increased \cite{ezhilan2013instabilities}. In other words, in dense suspensions, steric interactions between particles lead to nematic ordering. For this reason, we primarily choose to probe the effects of closure approximations as a function of the alignment strength $\zeta$, suppressing dependencies on $\alpha,\beta$. The dipole strength was set to $\alpha=-1$ and the particle density $\beta=0.8$, and the rotational and translation diffusion coefficients to $d_T = d_R = 0.05$.\\

\subsection{\textit{A priori} analysis of the learned closures}
In this section, we perform numerical experiments to compare the prediction accuracy of data-driven closure models based on an isotropic representation and component-wise learning. For the learning based on tensor-valued isotropic function representation, we use three input features based on the invariant quantities $\mathcal{I}_s(\mathbf{D},\mathbf{E})$ as described in Section~\ref{sec:invariant_rep}. For the component-wise learning, we provide the independent components $\{ d_{11}, d_{12}, d_{22}, e_{11}, e_{12}, e_{22}\}$ of each tensor $\mathbf{D},\mathbf{E}$ as the input to the neural network. For each value of the alignment strength parameter $\zeta$, we generated kinetic simulation data which was then used to train a neural network and make predictions. We use relative mean-square error (RMSE) as the performance metric:
\begin{equation}\label{eq:RMSE}
    \textrm{RMSE}(\zeta) =  \frac{\Vert \mathbf{S\!:\!T}(\zeta) - \widetilde{\mathbf{S\!:\!T}}(\zeta,\theta)\Vert_2^2}{\Vert\mathbf{S\!:\!T}(\zeta)  \Vert_2^2},
\end{equation}

\begin{figure}[htbp]
  \centering
  \includegraphics[width=6.25in]{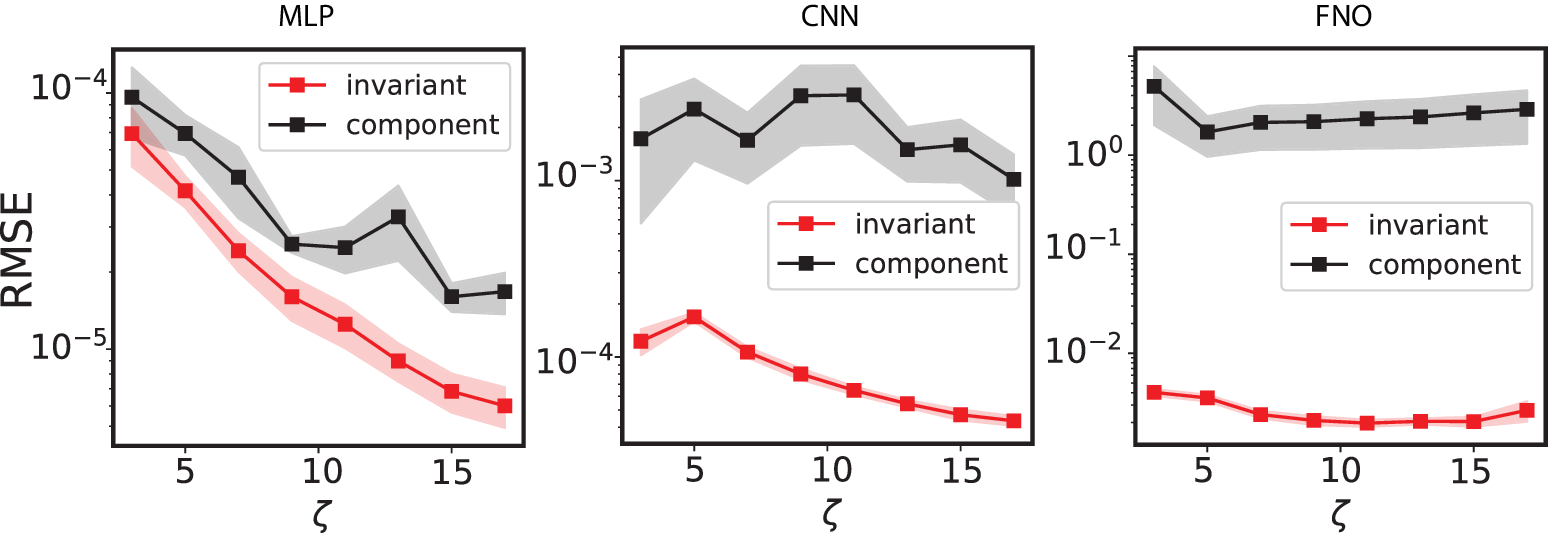}
  \caption{\textbf{Comparison between invariant based learning vs component-wise learning}: The Relative Mean Square Error (RMSE) for different network architectures MLP(left), CNN(middle), FNO(right). The shaded area around the line depicts one standard deviation from the mean. The neural networks have a comparable number of parameters, $~10000$, and architectures as shown in Table~\ref{network_architecture}.}
  \label{fig:comparison_networks}
\end{figure}
\noindent where $\mathbf{S\!:\!T}(\zeta)$ is reference data from kinetic theory generated with parameter $\zeta$ and $\widetilde{\mathbf{S\!:\!T}}$ is the prediction of the closure using neural networks. From Fig.~\ref{fig:comparison_networks}, we report that, irrespective of the network architecture, learning based on invariant representation leads to better performance in terms of prediction accuracy compared to component-wise learning. Invariant representation takes into consideration rotational symmetries during the learning process, which contributes to its superior performance. Interestingly, for a comparable number of network parameters and the same training data, going from point-wise learning (MLP) to learning based on global convolutions (FNO), the accuracy of the network predictions deteriorates. Especially, we see that FNO struggles to approximate the closure with component-wise learning and gains two orders of magnitude increase in prediction accuracy with the invariant-based representation for varying alignment strength $\zeta$. Thus, a representation based on tensor-valued isotropic function allows for learning accurate closures from a few time snapshots of kinetic theory simulations. Interestingly, from Fig.~\ref{fig:comparison_networks} it is also evident that learning the point-wise nonlinear map using MLP yields the best \textit{a-priori} predictions. The MLP achieves this performance without needing any additional local neighborhood information like in CNN or accounting for nonlocal effects through global convolutions as in the FNO architecture.

\subsection{Comparison with commonly used closures}
Several closures have been proposed in the context of the theory of passive and active liquid crystal polymers. In this study, we specifically compare our data-driven closure against the following established closure models: \\

\noindent \textit{Linear closure}: One of the simplest closures is derived from truncating the particle distribution function $\Psi$ on the basis of spherical harmonics and ignoring the coefficients of harmonics for degree greater than or equal to three \cite{saintillan2013active}. In two dimensions, the resulting linear closure can be expressed using the largest eigenvalue ($\mu_1$) of the normalized orientational order parameter $\mathbf{Q} = \mathbf{D}/c$ as,
\begin{equation}
    S_{1111}^{'} \approx \left( 3/8 + (\mu_1 - 1/2) \right) c,
\end{equation}
where $S_{ijkl}^{'} = \Omega_{im}\Omega_{jn}\Omega_{kp}\Omega_{lq}S_{mnpq}$ is the rotated fourth-moment tensor in the diagonal frame of $\mathbf{D}$, i.e. $\mathbf{D} = \mathbf{\Omega} \mathbf{D}^{'} \mathbf{\Omega}^T$. Here, $\mathbf{\Omega}$ is an orthonormal matrix and
$\mathbf{D}^{'} = \textrm{diag}\{ \mu_i\}_{i=1}^d$ is a diagonal matrix consisting of the ordered eigenvalues of $\mathbf{D}$. The full tensor $S_{ijkl}$ can be computed from $S_{1111}'$ using the trace identities $S_{1111}' + S_{1122}' = \mu_1$ and $S_{1122}' + S_{2222}' = c(1-\mu_1)$ and rotations.\\

\noindent \textit{Quadratic closure}: This \textit{ad hoc} closure proposed by Doi \cite{doi1981molecular} approximates the fourth moment $\mathbf{S}$ as the quadratic dependence of the second moment $\mathbf{D}$, i.e. $S_{1111}^{'} \approx \mu_1^2.$ Despite the \textit{ad hoc} nature of the quadratic closure, it is very commonly used in theories for passive and active liquid crystals \cite{edwards1990generalized, marenduzzo2007steady}.\\

\noindent \textit{Bingham closure}: Originally introduced by Chaubal and Leal \cite{chaubal1998closure} in the context of liquid crystal polymers, the Bingham closure models the particle distribution function as
\begin{equation}
    \mathbf{S}_\mathbf{B}[\mathbf{D}] = Z^{-1} \int_{\vert \mathbf{p}\vert=1} \mathbf{pppp} \: e^{\mathbf{B}[\mathbf{D}]:\mathbf{pp}} ~ d\mathbf{p}.
\end{equation}
This closure relies on solving the inverse problem of determining the parameters $\mathbf{B}$ and $Z$ such that following moment constraints are satisfied at each point in space,
\begin{equation*}
c(\mathbf{x},t) = \frac{1}{Z(\mathbf{x},t)} \int_{\vert \mathbf{p}\vert=1} e^{\mathbf{B}(\mathbf{x},t):\mathbf{pp}} d\mathbf{p}; \quad 
\mathbf{D}(\mathbf{x},t) = \frac{1}{Z(\mathbf{x},t)} \int_{\vert \mathbf{p}\vert=1} \mathbf{pp}e^{\mathbf{B}(\mathbf{x},t):\mathbf{pp}} d\mathbf{p}.
\end{equation*}
\noindent The recent work \cite{weady2022fast} proposed an efficient method for evaluating the Bingham closure using fast and accurate Chebyshev representation to reconstruct the fourth-moment tensor to near machine precision, with accuracy and efficiency that can be finely controlled. By solving the above inverse problem over the feasible range of $\mu_1 \in[1/2/1]$, this provides the following closure representation,
\begin{equation}
 S_{1111}^{'} \approx \sum_{m=0}^M c_m T_m(4\mu_1 - 3),
\end{equation}
where $T_m(\nu)$ is the $m^{th}$ Chebyshev polynomial with coefficient $c_m$. 
One can then calculate the closure $\mathbf{S\!:\!T}$ by performing a contraction in the diagonal basis of $\mathbf{D}$ and rotating back $\mathbf{S\!:\!T} = \mathbf{\Omega} \cdot \!\left(\mathbf{S^{'}\!:\!T^{'}} \right) \! \cdot \mathbf{\Omega}^\top$.\\

\begin{figure}[htbp]
  \centering
  \includegraphics[width=4.0in]{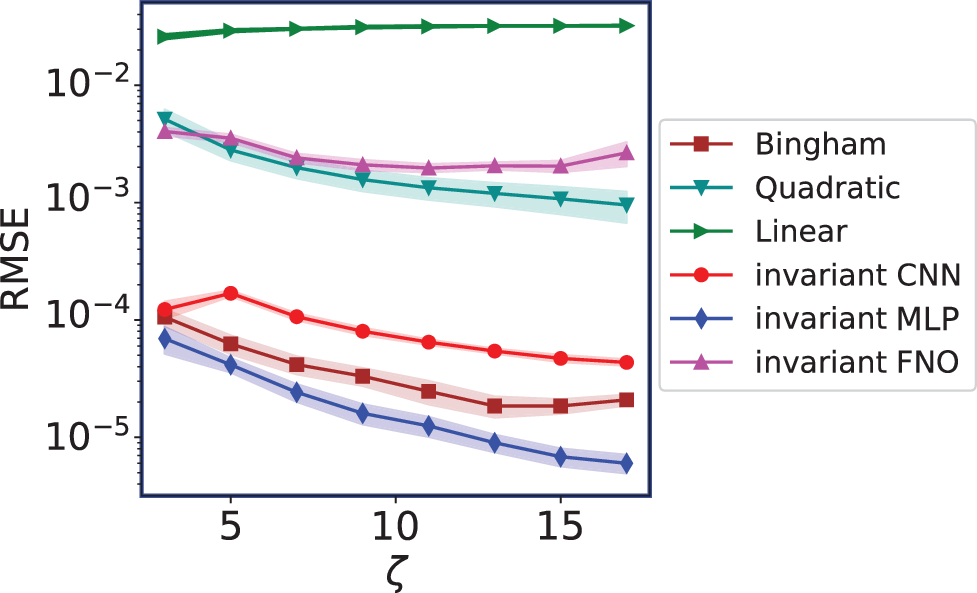}
  \caption{{\textbf{Comparison of existing closures with data-driven approximation:}} We compare the RMSE of the linear, quadratic and Bingham closures to those based on neural networks for different values of the alignment parameter $\zeta$. The CNN and MLP yield the highest accuracy and perform similarly to the Bingham closure, while the FNO is less accurate by two orders of magnitude.}
  \label{fig:compare_classics}
\end{figure}

\begin{figure}[htbp]
  \centering
  \includegraphics[width=6.5in]{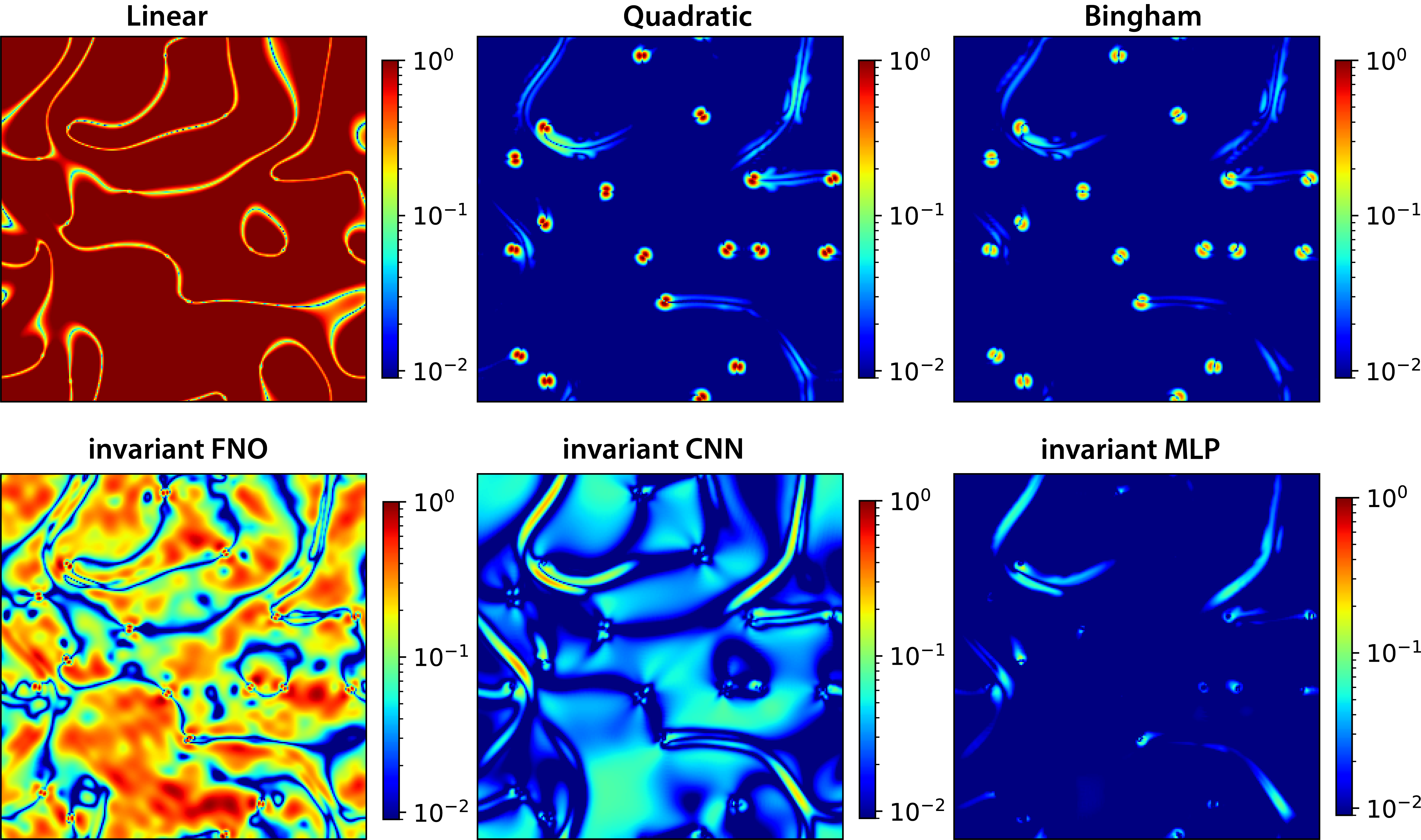}
  \caption{{\textbf{Comparing point-wise absolute error:}} The point-wise error across the spatial domain is shown, representing the predictions of a component of the tensor quantity $\widetilde{\mathbf{S\!:\!T}}$ using different closures for $\zeta=15$. Compared to other closures, the MLP produces the lowest error and yields accurate solutions near defects which other closures fail to capture.}
  \label{fig:compare_snapshots}
\end{figure}

\noindent In Fig.~\ref{fig:compare_classics}, we plot the RMSE for predictions based on the closures discussed above and data-driven closures discussed in the previous section. We note that neural networks based on point-wise (MLP) and patch-wise filter-based learning (CNN) with invariant features are able to reach prediction accuracy comparable to Bingham closure for different values of $\zeta$ (see Fig.~\ref{fig:compare_classics}). Also, the MLP and CNN architectures with invariant representation gain two orders of magnitude in prediction accuracy compared to the commonly used quadratic closure \cite{doi1981molecular}. This is also evident in Fig.~\ref{fig:compare_snapshots} where we show single time snapshots of the point-wise prediction error $\vert \mathbf{v} - \widetilde{\mathbf{v}}\vert$ using different closures for $\zeta=13$. On closer observation in Fig.~\ref{fig:compare_snapshots}, we show the closure approximation using linear, quadratic and FNO closures struggle to adequately resolve regions close to defects. Given that these rich topological structures are central to the dynamics, it is important to adequately resolves the features associated with them for accurate characterization of the spatiotemporal dynamics. 
\begin{figure}[htbp]
  \centering
  \includegraphics[width=6.0in]{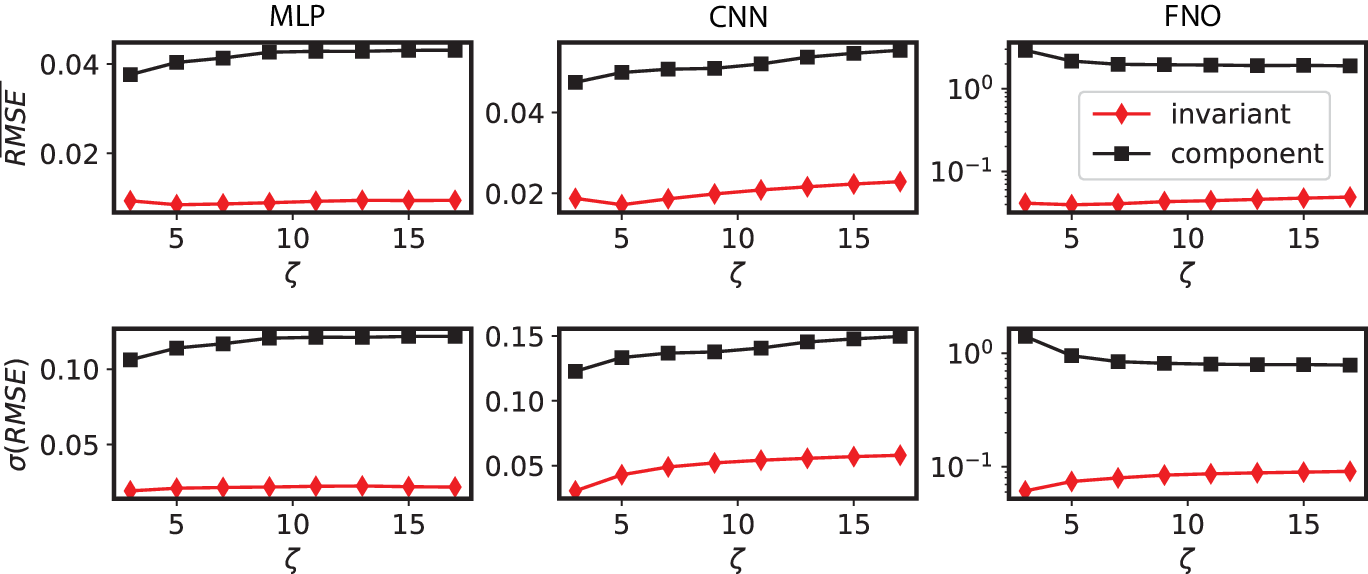}
  \caption{{\textbf{Extrapolation properties of invariant based learning (red) vs.\ component-wise learning (black):}} For each value of the parameter $\zeta$, we compare the mean (top row) and standard deviation (bottom row) of predictions made by neural networks trained using data generated from different values of $\zeta$.}
  \label{fig:extrapolation}
\end{figure}

Next, we test how different neural network architectures generalize to different parameter regimes. For this, we use a neural network trained with a specific value of the parameter $\zeta$ to make closure predictions at other values of the same parameter. To quantify the extrapolation accuracy, we use the following metrics:
\begin{equation*}
    \overline{\textrm{RMSE}}(\zeta) = \frac{1}{N_\zeta} \sum_j \textrm{RMSE}(\zeta,\theta_j), \quad  \sigma (\textrm{RMSE}) = \frac{1}{N_{\zeta}} \sum_j \left( \textrm{RMSE}(\zeta,\theta_j)  - \overline{\textrm{RMSE}}(\zeta)\right)^2.
\end{equation*}
Here, $\textrm{RMSE}(\zeta,\theta_j)$ is the relative mean square error associated with the prediction for parameter $\zeta$ using a neural network trained on the data generated with parameter $\zeta_j$, and $N_\zeta$ is the number of $\zeta$ values screened for prediction. In Fig.~\ref{fig:extrapolation}, we show the mean $\overline{\textrm{RMSE}}$ and standard deviation  $\sigma (\textrm{RMSE})$  associated with the predictions of neural networks trained at different values of $\zeta \in \{3,...,17\}$. Once again, we consistently observe that regardless of the architecture employed, neural networks trained with invariant representation exhibit better prediction accuracy for extrapolating to a value of $\zeta$ that differs from the one used to generate the training data, in comparison to component-wise learning. 

Overall, we have shown that embedding rotational equivariance through tensor-valued isotropic representation improves both the accuracy and extrapolations abilities of all neural architectures. However, comparing Figs.~\ref{fig:comparison_networks} and \ref{fig:extrapolation}, we observe a significant drop in the prediction accuracy of neural networks. This is attributed to the dramatic change in characteristics of the fluid flow and particle orientation as the alignment strength is varied, and the over-fitting nature of neural networks. In the next section, we discuss an active learning procedure for training a neural network to predict and generalize well across a region of parameter space.

\subsection{Active learning for exploring parameter space}\label{sec:activelearning}

In the previous section, we have shown performance metrics with predictions using neural networks trained at the same value of the parameter $\zeta$. However, the goal is not to use a different neural network when the model parameters are varied, but to be able to train a single network to accurately predict the closure term for any value of the parameter in a predefined region of interest, using training data from as few samples in the parameter space as possible \cite{arthurs2021active, fasel2022ensemble, settles2009active}. In order to achieve this, we employ an autonomous network training procedure based on active learning. This strategy involves exploring and soliciting training data that maximally inform the learning process, thereby improving the network's ability to accurately predict the closure term for a wide range of parameter values. \\


In this section, using a pool-based selective sampling strategy, we efficiently explore the parameter space characterizing dipole strength $\alpha$ and alignment strength $\zeta$ \cite{settles2009active}.
First, we choose the region $\mathcal{R}$ of interest in the parameter space over which we want to predict the closure terms. We discretize the region $\mathcal{R}$ with the grid $\mathcal{G}$ and generate kinetic simulation data at each grid point. Then, an autonomous active learning procedure is employed with the following steps: 1) Randomly choose a point on the grid and compile the training data. 2) Train the neural network on the available training data. 3) Use the trained network to make a prediction of the closure term $\widetilde{\mathbf{S\!:\!T}}$ at all points on the grid, regardless of where previous training data were generated. 4) Augment the training data with data generated from the point on the grid $\mathcal{G}$ where the largest RMSE is observed and return to step 2.\\

\begin{figure}[htbp]
  \centering
  \includegraphics[width=7in]{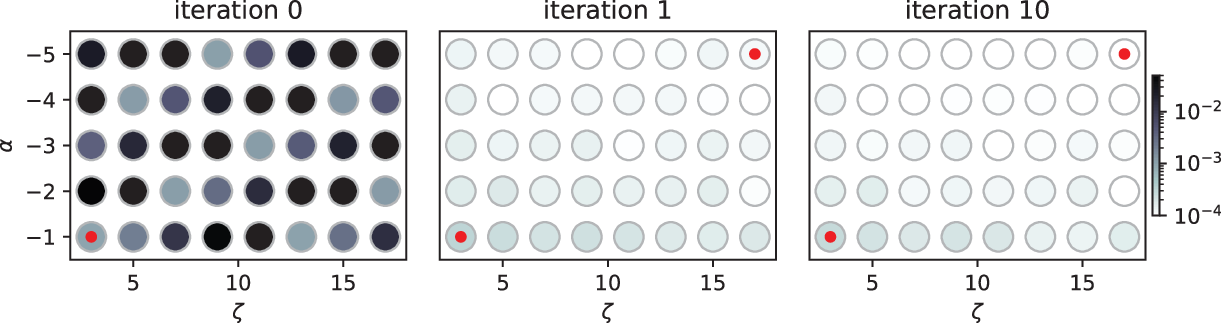}
  \caption{{\textbf{Parameter sampling through active learning:}} Different iterations (columns) of the training procedure employing pool-based selective actives strategy are shown. The red dots indicate the locations where kinetic simulation data is solicited and included in the training set. The color coding of the grid points represents the magnitude of the RMSE associated with the network's prediction.}
  \label{fig:al}
\end{figure}

\begin{figure}[htbp]
  \centering
  \includegraphics[width=7in]{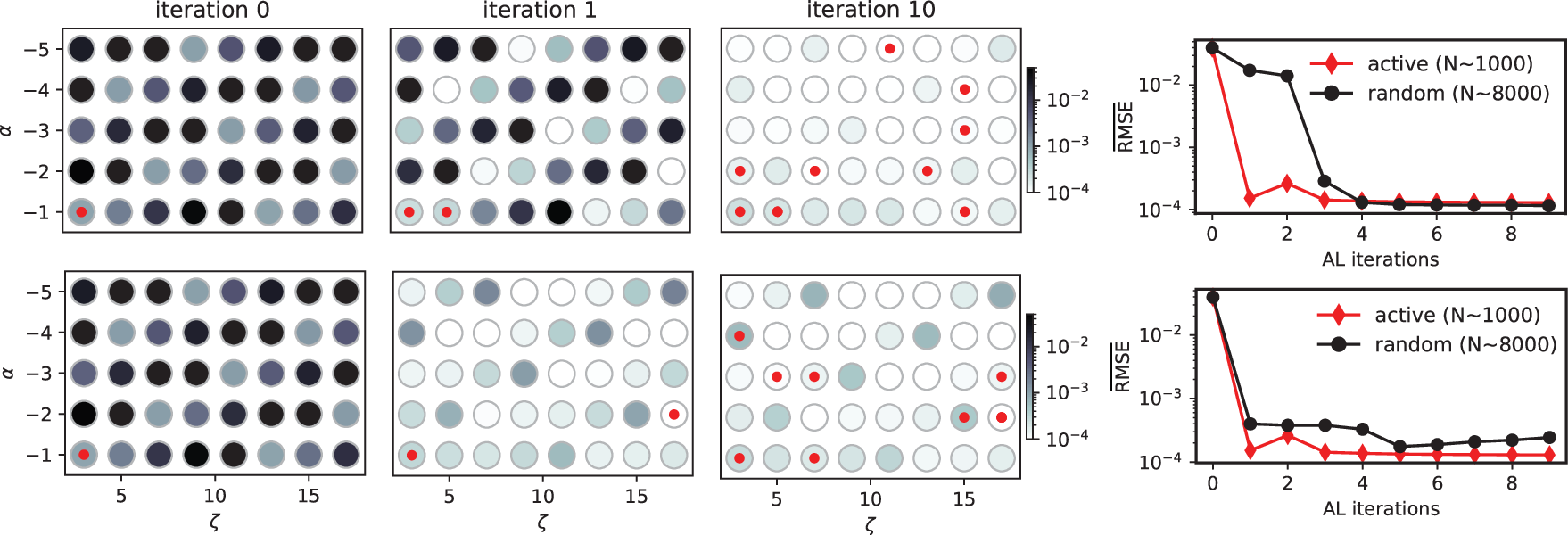}
  \caption{{\textbf{Random sampling:}} Different iterations of the training procedure based on random parameter selection are shown for two different seeds. The red dots indicate the locations where kinetic simulation data is selected and included in the training set. The color coding of the grid points represents the magnitude of the RMSE associated with the network's prediction. The last column presents the comparison of the $\overline{\textrm{RMSE}}$ across the entire grid during the training process contrasting the AL and random sampling strategies. }
  \label{fig:rl}
\end{figure}

\indent In Figure~\ref{fig:al}, we demonstrate the utility of the active learning procedure, where using just two points on the grid $\mathcal{G}$ we achieve better predictive performance than random sampling as shown in Figure~\ref{fig:rl}. The RMSE comparison in Figure~\ref{fig:rl} highlights the difference in performance between the active learning (AL) and random sampling strategies during the training process. As early as the second iteration of the AL procedure, the AL strategy achieves near-peak prediction performance showcasing its data efficiency and faster training process compared to random sampling. Also, the AL strategy in Figure~\ref{fig:al} demonstrates that selecting data points at the extrema of the parameter space (low and high) is sufficient to achieve accurate predictions across the entire parameter regime. This finding suggests that the flow and order parameters' characteristics in the intermediate parameter regime can be effectively captured by interpolating between the low and high values of the dipole and alignment strength. Unlike the active learning strategy, Figure~\ref{fig:rl} demonstrates that random sampling does not consistently reduce prediction errors across the entire grid.




\subsection{Learning stable \textit{a-posteriori} estimates through model prediction control}

In the previous sections, we demonstrated how neural networks trained with tensor isotropic representation can accurately approximate the closure terms. However, the supervised training used in the context of \textit{a-priori} analysis cannot express the long-term effects of the closure approximations and thus can lead to \textit{a-posteriori} predictions that are not in agreement with kinetic theory \cite{weady2022thermodynamically}. In particular, 
the models learned could be over-fitted and numerically unstable for long-term temporal predictions. In this section, we demonstrate how neural networks trained using supervised learning on static reference data can lead to unstable \textit{a-posteriori} predictions and training based on the deep integration of neural networks with a numerical solver is critical for resolving this issue.\\

\begin{figure}[htbp]
  \centering
  \includegraphics[width=5in]{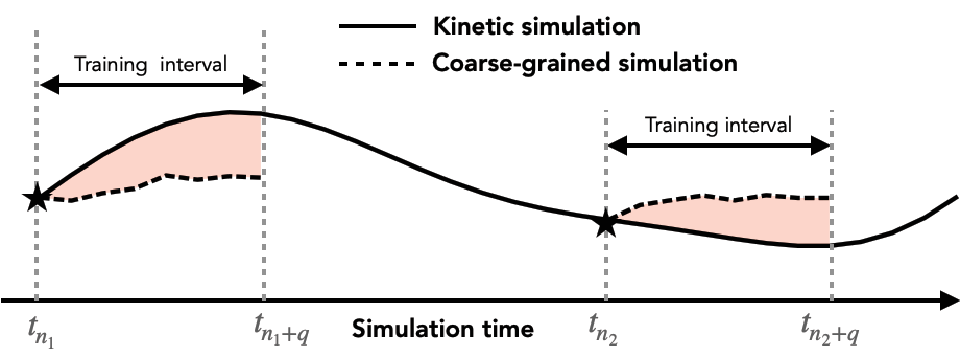}
  \caption{ {\textbf{Discrete Model predictive control through periodic injection:} Initial conditions are provided to the pseudo-spectral solver  at the injection points denoted by the set $\mathcal{I}^\star$ and illustrated in the figure by the star symbol. Coarse-grained predictions are made within the training horizon $\{t_{k},\hdots, t_{k+q} \}$. The shaded region illustrates the discrepancy between the predictions and the kinetic theory that is being minimized through the optimization procedure described in Eqs.~(\ref{mpc})-(\ref{predict_D}). }  }
  \label{fig:mpc_train}
\end{figure}

We propose an optimization procedure based on model predictive control (MPC) that allows us to close the gap between the usual \textit{a-priori} learning and a separate \textit{a-posteriori} testing.  In MPC, the closure approximation with the neural network is optimized during the training process such that the error between the predicted trajectory and the kinetic simulations over multiple time-steps is minimized (See Fig.~\ref{fig:mpc_train}). The method of model predictive control is akin to Neural ODEs \cite{chen2018neural, maddu2023, rudy2019deep}, where unstable models are penalized using time-stepping constraints. Alternatively, we can interpret the closure approximations as \textit{data-driven controls} that are learned to steer the system towards a desired target state set by kinetic simulations. The optimization formulation for model predictive is
\begin{align}
     \widetilde{\theta} = \argmin_{\theta } & \sum_{k \in \mathcal{I}^\star}\sum_{l=1}^{q} \nu_l \Bigg[\int_{\Omega_d} \left(\big \Vert  \mathbf{D}_\mathcal{K}^{k,l}(\mathbf{x}) \color{black} - \widetilde{\mathbf{D}}^{k,l}(\mathbf{x},\theta) \big \Vert_F^2 \right)d\mathbf{x}\Bigg] + \lambda_{wd} \sum_{i=0}^{N_l} \Vert \mathbf{W}_i \Vert_F^2, \label{mpc} \\
         \textrm{where} & \quad \widetilde{\mathbf{D}}^{k,l} = \mathcal{T}_{\mathbf{D}}^{l} \left( \widetilde{\mathbf{D}}^{k,l-1}, \widetilde{\mathbf{u}}^{k,l-1}, \mathcal{N}_\theta \left( \widetilde{\mathbf{D}}^{k,l-1},\widetilde{\mathbf{E}}^{k,l-1} \right) \right), \quad  \widetilde{\mathbf{D}}^{k,0} = \mathbf{D}_\mathcal{K}^{k,0}. \label{predict_D}
\end{align}

\noindent Here, the positive integer $q$ is referred to as the training time horizon, i.e. the number of time-integration steps considered during optimization, and $\theta$ represents the parameters of the neural networks that are being optimized. The scalars $\nu_l$ are exponentially decaying (over the training time horizon $q$)
weights that account for the accumulating prediction error \cite{rudy2019deep, maddu2023}. We optimize the trajectory over a small prediction window $q=2,3,4$ as the kinetic simulations have finite Lyapunov exponents. The propagator $\mathcal{T}_\mathbf{D}$ is the function that updates the orientation tensor $\mathbf{D}$ and the velocity $\mathbf{u}$, and encapsulates all the time-stepping formulae for $\mathbf{D}$ and the Stokes operator used to update $\mathbf{u}$. The neural network $\mathcal{N}_\theta(\cdot)$ takes as input the invariant quantities $\{\textrm{tr} \mathbf{D}^2, \textrm{tr} \mathbf{E}^2, \textrm{tr} \mathbf{DE} \}$ and outputs the closure approximation $\widetilde{\mathbf{S\!:\!T}}$ which is then passed through the propagator $\mathcal{T}_\mathbf{D}$ for prediction.
The prediction from the spectral solver, $\widetilde{\mathbf{D}}$, is then compared with the kinetic data, $\mathbf{D}_\mathcal{K}$. Due to the coupling between $\mathbf{D}$ and $\mathbf{u}$ through the Stokes equation, during optimization we compare errors in trajectories for only $\mathbf{D}$ as described in equation~(\ref{mpc}). We also impose a penalty on the weights of the network, $\mathbf{W}_{i=1,2,...,N_l}$, in order to prevent over-fitting with the penalization strength controlled by the constant $\lambda_{wd}$. We performed the \textit{a posteriori} analysis using the MLP architecture as it yielded the best prediction accuracy in \textit{a-priori} analysis.\\

We use similar MLP architectures employed in the previous section for \textit{a-priori} analysis. The neural network in the non-linear MPC is trained for $10000$ epochs with an initial learning rate $\eta=0.01$. The differentiable pseudo-spectral solver is integrated with time-step $\Delta t=0.01$ with initial conditions provided from the kinetic simulations. For fast and efficient optimization, we make short window predictions with initial conditions prescribed at predefined injection points along the time axis as shown in Figure \ref{fig:mpc_train}. This also allows for natural batching of the data with each batch corresponding to an injection point and its corresponding prediction window. Given that during training we only optimize over short training intervals, we leverage automatic differentiation to efficiently compute the gradients of the learned model for gradient-based nonlinear optimization.
Note for optimization and control over long-time windows the low-level auto-grad differentiation becomes computationally demanding and memory-intensive. In such scenarios, techniques like adjoint methods can be used to calculate derivatives efficiently, even for large simulations with thousands of
control parameters (neural network parameters) \cite{mcnamara2004fluid}.

\begin{figure}[htbp]
  \centering
  \includegraphics[width=6.5in]{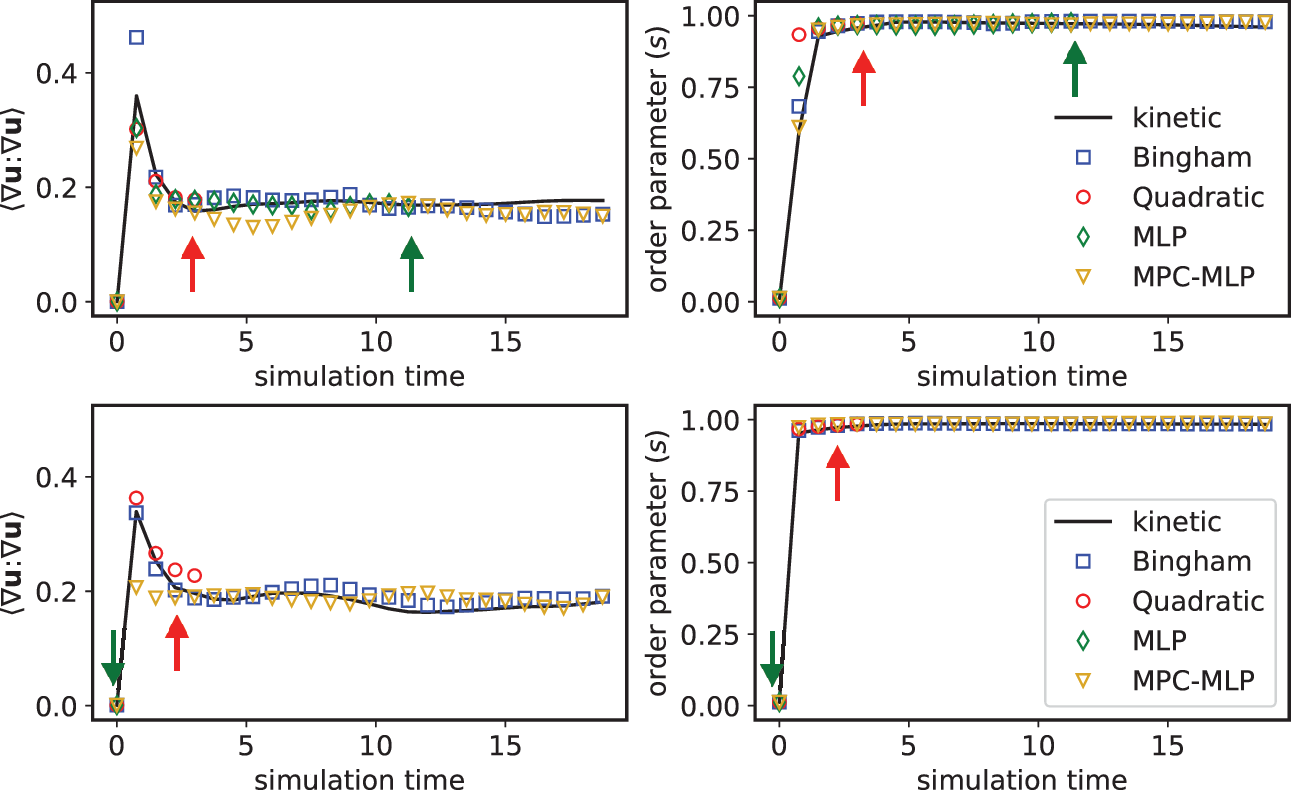}
  \caption{{\textbf{Comparison of mean velocity statistics $\langle \nabla \mathbf{u}\!: \!\nabla \mathbf{u}\rangle_V$ and scalar order parameter $(s)$ :}} The \textit{a-posteriori} tests are performed under different closure models with alignment strength  $\zeta=7$ (top row) and $\zeta=15$ (bottom row). Different colored symbols correspond to different closure models employed as described in the inset. The colored arrows (red: quadratic, green: MLP) represent the time point at which the coarse-grained simulations becomes unstable. The length of prediction window was set to $q=4$ to train neural networks based on MPC procedure. }
  \label{fig:compare_post}
\end{figure}

\begin{figure}[htbp]
  \centering
  \includegraphics[width=6.5in]{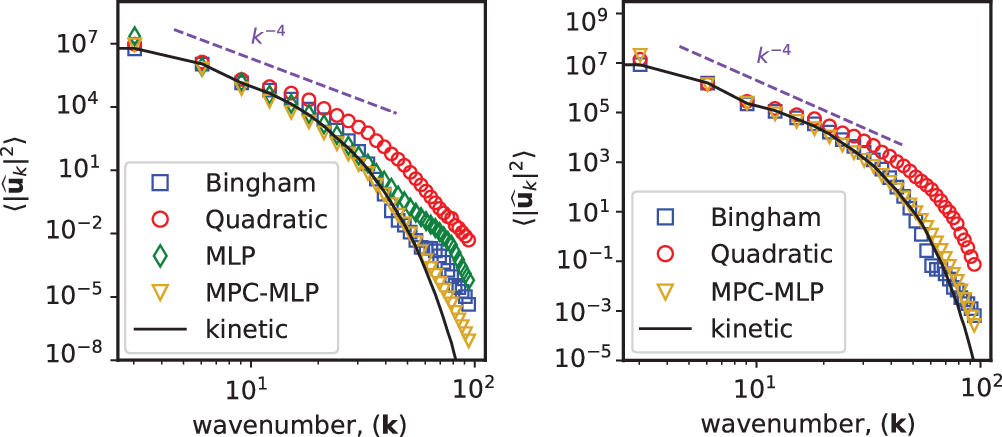}
  \caption{{\textbf{Power spectrum comparison:}} Velocity power spectrum calculated from coarse-grained simulations under different closure models for alignment strength $\zeta=7$ and $\zeta=15$.}
  \label{fig:compare_spec}
\end{figure}

\begin{figure}[htbp]
  \centering
  \includegraphics[width=3.0in]{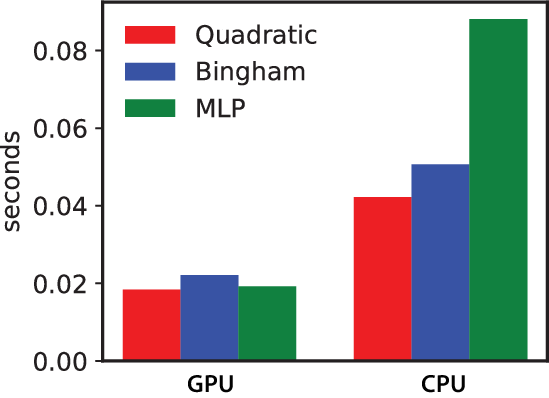}
  \caption{{\textbf{Computational cost:}} Cost per time-step of the coarse-grained simulations under different closure models. The GPU timings were measured on a Nvidia A100 graphics card and CPU timings were measured on an Intel(R) Xeon(R) Gold 6148 CPU @ 2.40GHz.}
  \label{fig:compare_cost}
\end{figure}

In Figure~\ref{fig:compare_post}, we compare, kinetic simulations with the coarse-grained simulations based on  mean statistics of the velocities and the scalar order parameter $ s(t)  = 2 \langle \mu_1(\mathbf{x},t) \rangle_{\mathbf{x}} -1 $. The angle brackets $\langle \cdot \rangle_{\mathbf{x}}$ denote averaging over the spatial domain. The predictions based on the Bingham closure and invariant MLP trained with the MPC procedure closely trace the trajectories of kinetic simulations, whereas predictions based on the quadratic closure and on the invariant MLP trained with supervised learning lead to unstable simulations. Figure \ref{fig:compare_post} illustrates that Model-Predictive Control clearly improves the temporal stability of the data-driven closures in comparison with models trained with supervised learning. This is also very much reflected in the velocity power spectrum comparison shown in Fig.~\ref{fig:compare_spec}, where the quadratic closure and MLP based on supervised learning do not adequately resolve the intermediate to large wavenumbers regime of the spectrum. We could not find any stable numerical results for the linear closure model for alignment strengths of  $\zeta=7$ and $\zeta=15$. In Figure~\ref{fig:compare_cost} we report the cost per time-step of the coarse-grained simulations under different closure models. On GPUs, the computational cost shows little variation across different closure models. Despite having  $\sim 10000$ parameters, the neural network (MLP) exhibits a similar evaluation time as the quadratic and Bingham closures.
However, when running on a CPU, the quadratic and Bingham closures outperform the MLP-based predictions in terms of speed.

\section{Parameter inference with nonlinear model predictive control}
In the previous sections, we analyzed the learned closures in the context of \textit{a-priori} estimates and stable and accurate \textit{a-posteriori} predictions. However, the utility of hydrodynamic theories is not just restricted to making qualitative predictions that resemble experiments but also can be used to infer important material properties of the system. Recent studies leverage machine learning for inferring activity fields from time snapshots of director fields \cite{colen2021machine,frishman2021learning}. In these studies, the problem of inferring material properties is recast as a computer vision problem, and a neural network is used in a pure black-box setting to output parameter values given a huge data library of director fields. In this section, we demonstrate how accurate coarse-grained descriptions and differentiable numerical solvers can conduct parameter inference conditioned directly on the available data. More specifically, we probe how different closure models affect the accuracy of the inferred parameters.\\

\begin{figure}[htbp]
  \centering
  \includegraphics[width=5.5in]{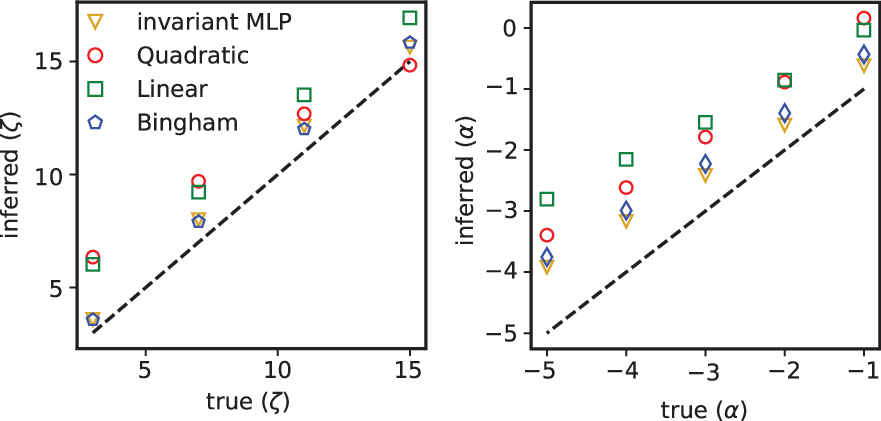}
  \caption{{\textbf{Parameter inference:}} Left: Alignment strength $\zeta$ inferred under different closures from data generated through kinetic theory with the dipole strength fixed at $\alpha=-1$ . Right: Inference of dipole strength $\alpha$ at fixed alignment strength $\zeta=3$.}
  \label{fig:compare_parameter}
\end{figure}

By using nonlinear MPC and a differentiable pseudospectral solver, we explore how accurately the coefficients of the kinetic theory can be inferred from simulations of the coarse-grained description (Eq.~\ref{coarse_grain_c},\ref{coarse_grain_D}) under a given closure model. The nonlinear MPC problem for inferring the alignment strength parameter $\zeta$ is
\begin{align}
      \widetilde{\Sigma} =& \argmin_{\Sigma}  \sum_{k \in  \mathcal{I}^\ast}\sum_{l=1}^{q} \nu_l \Bigg[ \int_{\Omega_d} \left(  \big \Vert  \mathbf{D}_\mathcal{K}^{k,l}(\mathbf{x}) \color{black} - \widetilde{\mathbf{D}}^{k,l}(\mathbf{x},\Sigma) \big \Vert_F^2 \right) d\mathbf{x} \Bigg] + \lambda_{wd} \sum_{i}^{N_l} \Vert \mathbf{W}_i \Vert_F^2,  \\    
    \text{where} & \quad \widetilde{\mathbf{D}}^{k,l} = \mathcal{T}_{\mathbf{D}}^l \left(\Sigma,  \widetilde{\mathbf{D}}^{k,l-1}, \widetilde{\mathbf{u}}^{k,l-1}, \mathcal{S}\left(\widetilde{\mathbf{D}}^{k,l-1},\widetilde{\mathbf{E}}^{k,l-1} \right)\right), \text{and } \widetilde{\mathbf{D}}^{k,0} = \mathbf{D}_\mathcal{K}^{k}.
\end{align}
Here, $\mathcal{S}$ is some prescribed closure model approximating $\mathbf{S\!:\!T}$ and $\Sigma$ is the parameter being inferred. As expected, Figure \ref{fig:compare_parameter} shows that both the Bingham closure and the data-driven closure lead to better parameter estimation. The MLP architecture was trained using the invariant input feature representation and the nonlinear MPC procedure discussed in the previous section. To be able to predict across different values of parameters, the neural network used for parameter estimation was trained using the active learning strategy described in section \ref{sec:activelearning}. We found the linear and quadratic closures encountered difficulties in parameter inference at low alignment strengths, which is in line with the \textit{a-priori} estimates shown in Figure \ref{fig:compare_classics}. In general, we found that estimating the dipole strength $\alpha$ at low alignment strength leads to poor parameter estimates. The relative error is close to $20\%$ for both Bingham and data-driven closure, $30\%$ for the quadratic closure, and $\approx 40\%$ for the linear closure. \\

The poor parameter estimates can be attributed to an ill-conditioned loss landscape potentially with multiple local minima associated with the nonconvex optimization problem. These issues are further exacerbated due to the reduced accuracy of the closure models at low values of alignment strength.
Future research should focus on addressing this challenge, possibly through advanced optimization techniques that can navigate the ill-conditioned landscape more efficiently, or by exploring alternate formulations that inherently lead to well-conditioned optimization problems \cite{frerix2021variational, gabrie2022adaptive}. This direction holds promise for improving parameter estimation and rendering the closures more accurate and reliable for inverse problems.



\section{Conclusion and Outlook}
We have presented a learning framework based on tensor-valued isotropic function representation of the closure terms arising in continuum models of active suspensions. We conduct both \textit{a-priori} and \textit{a-posteriori} analyses of the learned closures and provide a quantitative comparison with existing closures. Our comparison benchmarks clearly show that, with no additional cost, neural networks trained on rotationally invariant features have better prediction accuracy and generalizing power than those trained with tensor components as input features. Furthermore, we show that the learning framework based on invariant representation results in a two-order of magnitude increase in prediction accuracy when compared to widely employed closure models in passive and active liquid crystal theories. We also provide performance benchmarks between different neural network architectures with varying receptive and response fields. The MLP neural architectures based on point-wise learning between the input and the target yielded the best predictive performance, in comparison with CNN and FNO. In particular, for a comparable number of training parameters, the FNO architecture based on global convolutions performed poorly relative to MLP and CNN.
This demonstrated that nonlocal effects can be safely ignored to model the closure terms. 
To efficiently explore the parameter space, we use an active learning strategy, training a single neural network to predict across the parameter regime by interpolating between the low and high values of both the dipole coefficient and the alignment strength.
\\
\indent To bridge the gap between \textit{a-priori} analysis and \textit{a-posteriori} predictions, we developed a compute and memory efficient optimization procedure based on nonlinear model predictive control (MPC). By integrating neural networks with pseudo-spectral solvers, we were able to induce temporal stability prior to the learning problem. Within the discrete MPC framework, the neural network is trained to approximate the closure terms such that the discrepancy between the numerical solver's prediction and the kinetic theory is minimized within a prescribed training time horizon. In this way, we learn closures that can update the moments of the distribution function in a stable manner, which is not necessarily guaranteed from closures learned through supervised learning approaches with separate \textit{a-posteriori} testing. When run in inference mode, 
the coarse-grained simulations based on the learned models trained with nonlinear MPC remained stable, and accurately reproduced the mean velocity statistics, global order parameters, and velocity spectra of the kinetic theory. We leverage our differentiable pseudo-spectral solver with the data-driven closure to conduct parameter inference. Unsurprisingly, our analysis indicated a correlation between the accuracy of the closure model and the fidelity of the inferred parameters. Consequently, Bingham and data-driven closures yielded the most reliable estimates.\\
\indent In the present study, we only consider two-dimensional suspensions. However, extension of the formulation to three dimensions is straightforward. In three dimensions, new tensor bases and invariant quantities are added to the input features of the learning framework in its current form. However, generating three-dimensional kinetic simulations is computationally demanding and one would have to instead rely on highly resolved coarse-grained simulations, such as those based on the Bingham closure, as the ground truth. Future extensions where a differentiable solver that directly trains towards attaining \textit{a-posteriori} statistics like the scalar order-parameter or velocity spectra can be envisioned \cite{novati2021automating}.\\
\indent In this study, the focus was solely on the characterization of apolar suspensions and the representation of closure terms that emerge in their coarse-grained models. However, many active matter systems have polar order, such as microtubule and motor protein assemblies or collections of motile bacteria. Future work should extend the current learning framework to account for polarity, thereby describing a wider range of active matter systems. In summary, data-driven techniques provide a systematic approach to ensure consistency between coarse-grained macroscopic descriptions and the microscopic details of the underlying physical process.



\newpage
\appendixpage
\begin{appendices}

\section{Isotropic Tensor-valued functions in two dimensions}\label{derivation}
In this section, we derive the form of the second-order tensor-valued isotropic functions $\mathbf{S\!:\!D}$ and $\mathbf{S\!:\!E}$, which are functions of the tensor arguments $\mathbf{D}$ and $\mathbf{D,E}$, respectively. To do this, we first discuss the general representation of a scalar-valued isotropic function $f$ that has symmetric tensor-valued arguments $\{\mathbf{A}_i\},i=1,...,M$. The general representation theorem \cite{korsgaard1990representation} states that the isotropic function $f$ can be expressed in terms of the invariants of the functional basis of the argument tensors,
\begin{equation}\label{general_iso}
    f(\mathbf{A}_1,...,\mathbf{A}_M) = f(\mathcal{I}_s),
\end{equation}
where $\mathcal{I}_s, (s = 1, 2, . . . , S)$ are the invariants of the functional basis. The Cayley-Hamilton theorem \cite{pope1975more, korsgaard1990representation} can be used to deduce the minimal functional basis of an arbitrary set of tensors. In two-dimension, the functional basis is given as,
\begin{equation}\label{functional_basis}
    \{ \text{tr}{\mathbf{A}}_i, \text{tr}{\mathbf{A}}_i^2, \text{tr}{\mathbf{A}_i\mathbf{A}_j}\}_{i,j=1,2,...,M; \: i < j}
\end{equation}
To derive the expression for $\mathbf{S\!:\!E}=\mathbf{F}(\mathbf{D},\mathbf{E})$, we introduce a symmetric tensor $\mathbf{C}=\mathbf{C}^\top$ that forms a scalar-valued isotropic function $f$ through the inner product between the tensors $\mathbf{C}$ and $\mathbf{F}(\mathbf{D},\mathbf{E})$, 
\begin{equation}
    f(\mathbf{C},\mathbf{D,E}) = \textrm{tr}(\mathbf{C}\mathbf{F}(\mathbf{D,E})).
\end{equation}
It can be easily demonstrated that the function $f(\mathbf{C},\mathbf{D},\mathbf{E})$ is an isotropic function of its two symmetric tensor arguments due to the transformation invariance of the trace operator, as shown below:
\vspace{-0.5em}
\begin{align*}
    f(\bm{\Omega}\mathbf{C}\bm{\Omega}^\top,\bm{\Omega} \mathbf{D} \bm{\Omega}^\top,\bm{\Omega} \mathbf{E} \bm{\Omega}^\top) &=  \textrm{tr}(\bm{\Omega}\mathbf{C}\bm{\Omega}^\top\mathbf{F}(\bm{\Omega}\mathbf{D}\bm{\Omega}^\top,\bm{\Omega}\mathbf{E}\bm{\Omega}^\top)) \\
    &= \textrm{tr}(\bm{\Omega}\mathbf{C}\mathbf{F}(\mathbf{D,E})\bm{\Omega}^\top) \\
    &= \textrm{tr}(\mathbf{C}\mathbf{F}(\mathbf{D,E})) = f(\mathbf{C},\mathbf{D,E}).
\end{align*}
From Eqs.~(\ref{general_iso})-(\ref{functional_basis}), we see that any scalar-valued isotropic function $f$ can be represented in terms of the invariants of the functional basis of its argument tensors. The minimal functional basis associated with the argument tensors $\mathbf{C},\mathbf{D,E}$ are
\begin{equation*}
\{\textrm{tr}\mathbf{D},\textrm{tr}\mathbf{C}, \textrm{tr}\mathbf{E},\textrm{tr}\mathbf{C^2},\textrm{tr}\mathbf{D^2},\textrm{tr}\mathbf{E^2},\textrm{tr}\mathbf{CD}, \textrm{tr}\mathbf{CE}, \textrm{tr}\mathbf{DE} \}.
\end{equation*}
\noindent Since, by construction, $f(\mathbf{C},\mathbf{D,E})$ is linear in the tensor argument $\mathbf{C}$, only the invariants $\{ \textrm{tr}\mathbf{C}, \textrm{tr}\mathbf{CD},\textrm{tr}\mathbf{CE} \}$ are considered. This restricts the form of the scalar function $f$ to:
\begin{equation}
    f(\mathbf{C,D,E}) = \textrm{tr}(\mathbf{C}\mathbf{F(D,E)}) = \textrm{tr}\big [  \mathbf{C} \left( \kappa_0 \mathbf{I} + \kappa_1 \mathbf{D}+\kappa_2 \mathbf{E}\right)\big ],
\end{equation}
where $\kappa_0, \kappa_1, \kappa_2$ are scalar functions of the invariants of $\mathbf{D,E}$. Therefore, the final form of a symmetric isotropic tensor-valued function of a symmetric tensor is
\begin{equation*}
    \mathbf{F}(\mathbf{D,E}) =  \mathbf{S\!:\!E} = \kappa_0 \mathbf{I} + \kappa_1 \mathbf{D} + \kappa_2 \mathbf{E}.
\end{equation*}
\noindent The expansion for the tensor-valued function $\mathbf{S\!:\!D}$ can be derived through a similar procedure by omitting the dependence on $\mathbf{E}$.

\newpage

\section{Hyper-parameters of neural network architectures}\label{appendixA}

\begin{table}[htbp]
  \centering
\begin{tabular}{ |c|c|c|c|c| } 
 \hline
  & $\#$ input channels &  $\#$layers & $\#$ filters/nodes & $\#$ output channels \\ 
 CNN invariant & 3 & 6 & 16 & 5\\ 
 CNN component & 6 & 6 & 16 & 6\\ 
 \hline
   & $\#$ input channels &  $\#$layers & $\#$ modes/width & $\#$ output channels \\ 
  FNO invariant & 3 & 3 & 16/2 & 5\\ 
  FNO component & 6 & 3 & 14/2 & 6\\ 
 \hline
   & input-size &  $\#$layers & $\#$ nodes & output-size \\ 
 MLP invariant & 3 & 5 & 50 & 5\\ 
  MLP component & 6 & 5 & 50 & 6\\ 
  \hline
\end{tabular}
\caption{Hyper-parameters of the different network architectures used for \textit{a-priori} and \textit{a-posteriori} analysis. In CNN architectures a filter of size $3\times3$ was used.} \label{network_architecture}
\end{table}

\end{appendices}

\newpage
\bibliographystyle{unsrt}
\bibliography{main}

\end{document}